\documentclass[12pt]{article}
\usepackage{eurosym}
\usepackage{graphicx}
\DeclareGraphicsExtensions{.eps,.pdf,.png,.jpg}
\usepackage[utf8]{inputenc}
\usepackage[T1]{fontenc}
\usepackage{indentfirst}
\usepackage[margin=0.8in,nomarginpar]{geometry}
\usepackage[final]{hyperref}
\usepackage{amsmath}
\usepackage{hyperref}
\usepackage{cite}
\usepackage{subcaption}
\usepackage{caption}
\usepackage{amssymb}
\usepackage{multirow}
\usepackage{color}
\newcommand{\eps}{\epsilon}

\newcommand{\be}{\begin{equation}}
\newcommand{\ee}{\end{equation}}
\newcommand{\ba}{\begin{array}}
\newcommand{\ea}{\end{array}}

\hypersetup{
colorlinks=true,
linkcolor=blue,
citecolor=blue,
filecolor=magenta,
urlcolor=blue
}

\begin{document}
\title{Universal Thermodynamics of Dunkl-Deformed Bose Gases: From Power-Law Traps to Physical Bounds}
%Dunkl-Deformed Thermodynamics of Bose %Gases in D-Dimensional Power-Law Traps

\author{M. Medani\thanks{%
m.medani@univ-chlef.dz} ,
%\\
%EndAName
%Laboratory for Theoretical Physics and Material Physics Faculty of Exact \\
%Sciences and Informatics, Hassiba Benbouali University of Chlef, Algeria.
%\and
M. Benarous\thanks{%
m.benarous@univ-chlef.dz},  
%\\
%EndAName
%Laboratory for Theoretical Physics and Material Physics Faculty of Exact \\
%Sciences and Informatics, Hassiba Benbouali University of Chlef, Algeria.\and 
A. Hocine 
\thanks{%
ah.hocine@univ-chlef.dz} ,
%\\
%EndAName
%Laboratory for Theoretical Physics and Material Physics Faculty of Exact \\
%Sciences and Informatics, Hassiba Benbouali University of Chlef, Algeria. 
%\and 
and F. Merabtine%
\thanks{%
f.merabtine@univ-chlef.dz} \\
%EndAName
Laboratory for Theoretical Physics and Material Physics, Faculty of Exact \\
Sciences and Informatics, Hassiba Benbouali University of Chlef, Algeria.
%\and B. Hamil\thanks{%
%hamilbilel@gmail.com} \\
%EndAName
%Laboratoire de Physique Math\'{e}matique et Subatomique,\\
%Facult\'{e} des Sciences Exactes, Universit\'{e} Constantine 1, Constantine,
%Algeria. 
%\and B. C. L\"{u}tf\"{u}o\u{g}lu\thanks{%
%bekir.lutfuoglu@uhk.cz (corresponding author) } \\
%EndAName
%Department of Physics, University of Hradec Kr\'{a}lov\'{e},\\
%Rokitansk\'{e}ho 62, 500 03 Hradec Kr\'{a}lov\'{e}, Czechia. 
}
\date{}
\maketitle

\begin{abstract}
\noindent
We study an ideal Bose gas confined by a $D$-dimensional power-law potential within the framework of the Dunkl formalism. By analyzing the combined effects of spatial dimensionality and trap geometry, we derive universal expressions for the thermodynamic quantities, which depend solely on a single parameter. This reduction reveals the existence of universality classes that apply to any power-law potential, regardless of its specific form.

\noindent
Furthermore, we demonstrate that the internal consistency of the Dunkl formalism requires the Wigner parameter to lie within the interval $[0, 2]$. This finding extends previous results obtained for harmonically trapped ideal Bose gases and establishes that these bounds hold for arbitrary regular potentials in any dimension.
\end{abstract}


\section{Introduction}

A particularly interesting extension of standard quantum mechanics has emerged through the Dunkl formalism \cite{Dunkl1}. This approach enables the exploration of deformed quantum systems that go beyond the conventional framework, capturing subtle physical effects such as modified symmetries or alterations in phase-space structure.

The Dunkl formalism modifies the Heisenberg algebra of position and momentum operators by introducing a deformed commutation relation\cite{chung}:
\begin{equation}
[ \hat{x}, \hat{p}] = i \hbar \left(1 + (\nu-1) \hat{R}\right),
\end{equation}
where $\hat{R}$ denotes the reflection operator: $\hat{R}f(x)=f(-x)$.
%\begin{equation}
%\hat{R}=\left(
%-1\right)^{x\frac{\partial}{\partial x}}, \quad
%{\rm with}\quad 
%\end{equation}
The parameter $\nu$ (more precisely, $\frac{\nu - 1}{2}$) is commonly referred to as the Wigner parameter. It is immediately apparent that, for this formalism to be meaningful, $\nu$ must be nonnegative.

This formalism has found applications across various fields of physics \cite{G1, phys1, G4, G3, Sargolzaeipor, phys2, Mota1, phys3, Kim, phys4, HHS, Merad, Mota2, Mota3, Bilel1, Bilel2, Bilel5, phys5, Bilel3, phys6} and mathematics \cite{kakei, Lapointe, Mik1, math1, Dunkl2, bie, math2, math3}, yielding parity-dependent solutions to a range of intriguing problems. More recently, the Dunkl formalism has been extensively applied in statistical physics, where deformation effects have been shown to significantly modify the partition function and, consequently, the resulting thermodynamic properties \cite{phys1}.
Most notably, the latter work established an equivalence between non-interacting systems within the Dunkl-deformed framework and interacting systems in the conventional (undeformed) formulation. This result paves the way for interpreting interactions as manifestations of algebraic deformations in the underlying mathematical structure.

In this context, a number of recent studies have examined how the Dunkl-deformed algebra influences the statistical and thermodynamic properties of ideal Bose gases undergoing Bose-Einstein condensation (BEC) \cite{phys5, HO24, phys6, bm25}. This line of research is driven by the central role that degenerate Bose gases, especially those confined in external potentials, play in modern physics—particularly in the study of macroscopic quantum phenomena such as BEC \cite{pitaevskii2003}. Additionally, these systems offer a useful framework for exploring how various factors affect their physical behavior. Indeed, the behavior of a Bose gas is highly sensitive to the dimensionality of the system. For example, in two dimensions, Bose-Einstein condensation (BEC) in an ideal Bose gas can occur only when a trapping potential is present. Furthermore, the specific form of the confining potential—whether harmonic, quartic, or otherwise—has a direct impact on both the statistical properties (such as the density of states and particle distribution) and the thermodynamic characteristics of the system, including phase transitions, critical temperature, and heat capacity.

The homogeneous ideal Bose gas in three dimensions was revisited in Ref.\cite{phys5}, where the impact of deformed algebra on the system was explored. Subsequently, Refs.\cite{HO24, phys6} shifted focus to trapped ideal gases: the former considered a 3D harmonic trap, while the latter studied power-law traps in one and two dimensions. These studies reported significant effects of the deformation on key thermodynamic properties such as the condensate fraction, critical temperature, and heat capacity. More recently, in Ref.\cite{bm25}, the analysis was extended to harmonically trapped ideal Bose gases in arbitrary dimensions. This work not only generalized the results of \cite{HO24}, but also revealed the existence of an upper bound on the Wigner parameter ($\nu<2$), in addition to the previously known lower bound ($\nu>0$), required for the system to exhibit a proper classical limit. A key conclusion of this study is that a deformed ideal Bose gas can represent a physically meaningful system only if the deformation parameter lies within this range. This naturally raises the important question of whether such bounds and constraints persist for arbitrary trapping potentials in any dimension. Given that Bose-Einstein condensation has now been experimentally achieved in low-dimensional systems with a variety of trapping configurations \cite{explowerd1, explowerd2, explowerd3, explowerd4, explowerd5, explowerd6, explowerd7, explowerd8, explowerd9}, it is pertinent to explore the extent to which these results on deformed statistics remain applicable.

The aim of this paper is to address this question by analyzing the implications of the formalism for general power-law trapping potentials in $D-$dimensional systems, highlighting the influence of both geometry and dimensionality.

The main results of our investigation can be summarized as follows:

First, we demonstrate that an ideal Bose gas confined by a potential of the form $r^{\eta}$ (where $\eta$ may be negative) is dynamically and statistically characterized by a single parameter, denoted by $s$. This parameter arises from a specific combination of the spatial dimension $D$ and the potential exponent $\eta$, and it defines universality classes that encompass various trapping configurations across different dimensions.

Second, we show that Bose-Einstein condensation in such systems corresponds to a second-order phase transition, typically accompanied by a discontinuity in the heat capacity—except in the special case $s=2$, where the transition is continuous. Notably, this includes the two-dimensional harmonic oscillator.

Finally, by analyzing the high-temperature limit, we find that thermodynamic quantities exhibit distinct classical behavior depending on the value of the deformation parameter $\nu$. This leads to the identification of an upper bound on $\nu$, required to ensure physical consistency. These findings extend and refine our earlier results in Ref.\cite{bm25}, providing a complete characterization of the range of $\nu$ for which Bose-Einstein condensation in deformed algebra remains physically meaningful.

The structure of the paper is as follows. In Section 2, we extend the analysis of power-law trapping potentials from Ref.\cite{phys6} to $D-$dimensional systems within the Dunkl formalism. In this framework, we show that the energy dependence of the density of states is governed by a nontrivial combination of the spatial dimension and the trap exponent, encapsulated in a single parameter $s$. This parameter plays a fundamental role in the thermodynamic properties of the system, allowing us to predict that Bose-Einstein condensation can only occur for $s>1$, independently of the deformation parameter $\nu$. Moreover, $s$ defines universality classes that group together different combinations of dimensions and trapping potentials exhibiting the same thermodynamic behavior. This universality will be demonstrated through explicit calculations of the transition temperature and the condensate fraction.

In Section 3, we examine the degenerate regime, showing that BEC is only possible if the deformation parameter $\nu$ satisfies a necessary consistency condition. We also explore the critical regime near the transition, where both the universality parameter $s$ and the deformation $\nu$ control the system's behavior. In this regime, we find that the heat capacity at constant volume exhibits the characteristic $\lambda-$point behavior, indicative of a second-order phase transition.

Section 4 is devoted to the high-temperature regime. Here, we show that the thermodynamic functions approach different classical limits depending on the value of $\nu$, which leads us to establish an upper bound on $\nu$ to ensure physically consistent classical behavior. Finally, we conclude with a summary of our main results and discuss possible extensions to more complex quantum systems, with the aim of uncovering new roles played by the Wigner parameter.

\section{The power-law potential and its universality classes}

In this section, we extend the analysis of \cite{phys6} to the $D-$dimensional case. Consider an ideal Bose gas composed of $N$ neutral atoms of mass $m$ confined by a $D$-dimensional power-law trap of the form
\begin{equation}
V(x)=V_0\left(\frac{|x|}{a}\right)^{\eta}
\end{equation}
where $x$ is the $D-$dimensional position vector, $V_0$ the height of the trap, $a$ its range and $\eta$ is a parameter describing its shape. For instance, for $\eta=2$ and $\eta=4$, one recovers the isotropic harmonic oscillator\cite{gro1,HO24} and the quartic potential\cite{gautam08}. Taking the limit $\eta\to -\infty$ corresponds to recovering the case of the homogeneous ideal Bose gas. Moreover, interpreting $V(x)$ as a representative term in the power series expansion of a broader class of regular potentials renders the limit $\eta\to +\infty$ physically meaningful. This limiting case will be utilized in our final analysis. 

To compute the statistical properties, it can be shown \cite{phys1, phys5} that the deformed partition function takes the form  
\begin{equation}
Z=\prod\limits_{i=1}^{\infty}
\frac{1+\left(z e^{-\beta
\eps _{i}}\right)^{\nu}}{1-\left(z e^{-\beta \eps _{i}}\right)^{2}},  \label{pf}
\end{equation}  
where $z=e^{\beta \mu}$ is the fugacity of the gas, $\beta =1/k_{B}T$ is Boltzmann's factor, and $\nu$ is the Wigner (or deformation) parameter. One readily notes that the limit $ \nu = 1 $ leads back to the well-known partition function of a  non-deformed ideal Bose gas. In equation (\ref{pf}), $ \eps_i $ denotes the single-particle energy of state "$i$".

The total particle number and internal energy can be obtained using the standard thermodynamic relations:  
\begin{equation}
N=z\frac{\partial\log Z }{\partial z }\bigg|_{\beta,V},  
\qquad  
U=-\frac{\partial \log Z}{\partial \beta}\bigg|_{z,V},  \label{UN}
\end{equation}  
which yield  
\begin{equation}
N=\sum_{i=0}^{\infty}n_{\nu}(\eps_i), \qquad
U=\sum_{i=0}^{\infty}\eps_i n_{\nu}(\eps_i),
\label{NU}
\end{equation}  
where  
\begin{equation}
n_{\nu}(\eps_i)=\frac{2}{\left(z^{-1}e^{\beta\eps _{i}}\right)^{2}-1}+
\frac{\nu}{\left(z^{-1}e^{\beta\eps_{i}}\right)^{\nu}+1},
\label{distrib}
\end{equation}  
is the deformed Bose-Einstein distribution.

Except for some very special cases ($\eta=2, 4$), the discrete sums (\ref{NU}) are quite hard to compute. An alternative approach is to assume a continuous spectrum
$\eps=p^2/2m+V(x)$
and introduce the density of states
\begin{equation}
\rho(\eps)=\int\, \frac{d^Dp\,d^Dx}{(2\pi\hbar)^D}\delta (\eps-\eps (p,x)).
\label{ro}
\end{equation}
%This is a kind of semiclassical approximation, where the spacing between energy levels $\epsilon$ is small, the density of state writes 
The total number of atoms, and the internal energy then write
%\begin{equation}
%N_{0} =\frac{2}{z^{-2}-1}+\frac{\nu}%{z^{-\nu}+1},
%\end{equation}
%\begin{equation}
%N=2\int_{0}^{\infty }\frac{\rho (\eps) d\eps}{\left(z^{-1}e^{\beta\eps}\right)^{-2}-1}+\nu\int_{0}^{\infty }\frac{\rho(\eps) d\eps}{
%\left(z^{-1}e^{\beta\eps}\right)^{-\nu}+1} .  
%\label{uuu}
%\end{equation}
\begin{equation}
N=\int_{0}^{\infty }n_{\nu}(\eps)\rho (\eps) d\eps , \qquad
U=\int_{0}^{\infty }n_{\nu}(\eps)\eps\rho (\eps) d\eps .  
\label{uuu}
\end{equation}
Carrying out the integration over $p$ in (\ref{ro}), and then over $x$, one readily gets
\begin{equation}
\rho(\eps)= A\,\eps^{D\left(\frac{1}{\eta}+\frac{1}{2}\right)-1},
\label{roeps}
\end{equation}
where
\begin{equation}
A=\frac{2}{\eta}\frac{\Gamma(D/\eta)}{\Gamma(D/2)\Gamma(D\left(\frac{1}{\eta}+\frac{1}{2}\right))}\left(\frac{ma^2}{2\hbar^2}\right)^{D/2}
\left(\frac{1}{V_0}\right)^{D/\eta}  
\label{ampl}
\end{equation}
is a constant independent of energy and will not appear in the final expressions. Most importantly, we observe that the $\eps$-dependence of $\rho(\eps)$ is governed by the parameter
$$
s = D\left(\frac{1}{\eta} + \frac{1}{2}\right),
$$
which captures the combined effects of the system's dimensionality and the shape of the confining potential. Specifically, $\rho(\eps)$ increases with $\eps$ when $s > 1$, and decreases when $s < 1$. The special case $s = 1$ corresponds to an energy-independent density of states, reminiscent of the one-dimensional harmonic oscillator or the two-dimensional ideal gas. Notably, the limits $\eta\to\pm\infty$ are well-defined and lead to $s\to D/2$.

These three regimes—$s > 1$, $s = 1$, and $s < 1$—map to different ranges of the potential exponent $\eta$, depending on the space dimension $D$. For example, the case $s > 1$ leads to $0 < \eta < 2$ for $D = 1$, $\eta > 0$ for $D = 2$, and $\eta > 0$ or $\eta < -6$ for $D = 3$. This last case includes the familiar harmonic oscillator or molecular potentials of the form $1/r^{|\eta|}$ (including the Lennard‐Jones potential).

\begin{figure}[h]
    \centering
    \includegraphics[width=0.5\linewidth]{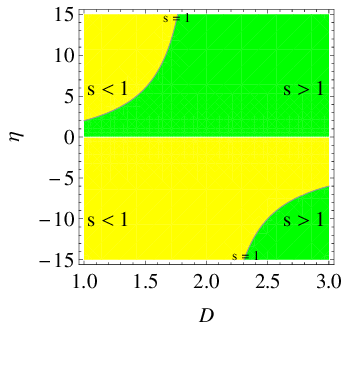} 
    % ou .pdf selon le format utilisé
    \caption{Universality classes characterized by the parameter $s$ as a function of spatial dimension $D$ and potential exponent $\eta$. Green regions correspond to parameter combinations where true Bose-Einstein condensation occurs ($s>1$), while yellow regions indicate regimes where quasi-condensation is expected ($s< 1$). The boundary at $s=1$ separates these distinct phases of quantum degeneracy. See text below.}
    \label{fig:figure4}
\end{figure}

Figure \ref{fig:figure4} illustrates the universality classes characterized by the parameter $s$ as a function of spatial dimension $D$ and potential exponent $\eta$. The regime $s<1$, encompasses physically relevant scenarios including repulsive power-law potentials ($\eta < 0$) in all dimensions, as well as steep attractive potentials with $\eta > 2$ in one dimension. Notably, three-dimensional systems exhibit this behavior for inverse power-law potentials in the range $-6 < \eta < 0$, which may be experimentally accessible through engineered optical potentials or modified inter-particle interactions.

The critical behavior of the density of states is determined by the parameter
$s$: for $s \leq 1$, the ground state maintains finite weight in the continuum limit, whereas for $s > 1$, this weight vanishes. This distinction has fundamental consequences for the possibility of Bose-Einstein condensation.

When the density of states $\rho(\varepsilon) \propto \varepsilon^{s-1}$ does not vanish sufficiently rapidly as $\varepsilon \to 0$
(i.e., $\leq 1$), the ground state becomes effectively invisible in the continuum approximation. Under these conditions, the discrete ground state cannot sustain macroscopic occupation relative to the dense manifold of low-energy excited states, precluding true Bose-Einstein condensation. This regime is instead characterized by quasi-condensation.
%, where density fluctuations are suppressed but phase coherence remains finite-ranged.

For $s > 1$, genuine Bose-Einstein condensation becomes possible as the ground state weight vanishes in the continuum limit, allowing for its explicit separation from the excited state spectrum. To ensure proper treatment of the condensate physics, we must account for the ground state population independently before applying the continuum approximation to the excited states.

We now validate this theoretical framework by computing the particle distribution while explicitly separating ground state and excited state contributions. Assuming $s > 1$ to ensure condensation is possible, we employ Eqs.~(\ref{NU}) and (\ref{roeps}) to obtain:

%The key aspect of the state density behavior is that for $s\le 1$, the ground state retains a non-zero weight, whereas for $s > 1$, its weight vanishes. As a result, the ground state is effectively lost when transitioning to the continuous limit. Therefore, one does not expect true Bose-Einstein condensation to occur for $s\le 1$. Given the fundamental role of the ground state in Bose-Einstein condensation, this issue must be addressed by explicitly separating out the number of atoms in the ground state before taking the continuous limit. 
%\textcolor{red}{Therefore, one does not expect true Bose-Einstein condensation to occur for $s\le 1$; instead, this regime is typically associated with quasi-condensation, where the density fluctuations are suppressed but the phase still fluctuates (quasi long-range order) \cite{psw2000}.}
%For s ≤ 1, the density of states doesn't vanish sufficiently fast near ε = 0, preventing the formation of a truly macroscopic ground state occupation. The system exhibits quasi-condensation with suppressed density fluctuations but persistent phase fluctuations, rather than the coherent, phase-stable condensate characteristic of true BEC.
%To validate these predictions—and simultaneously extend the analysis to the deformed case—we begin by computing the number of excited atoms, explicitly separating out the ground state population under the assumption that $s > 1$. Using equations (\ref{NU}) and (\ref{roeps}), we obtain:
\begin{equation}
N_{0} =\frac{2}{z^{-2}-1}+\frac{\nu}{z^{-\nu}+1},
\label{ss}
\end{equation}
\begin{equation}
N_{ex} =A \frac{\Gamma(s)}{\beta^s} g_s(z,\nu) ,  
\label{s}
\end{equation}
where $g_s(z,\nu)$ is a "generalized" Bose function
\begin{equation}
g_s(z,\nu) =2^{1-s}g_s(z^2)-\nu^{1-s}g_s(-z^{\nu}) ,  \label{GS}
\end{equation}
defined as a combination of the standard Bose (Polylogarithmic) functions $g_s(z)$:
\begin{equation}
g_s(z)=\frac{1}{\Gamma (s)}\int_{0}^{\infty }\frac{x^{s-1}}{z^{-1}e^{x}-1}dx.
\end{equation}
One notices in particular that $g_s(z,\nu=1)=g_s(z)$, leading to the standard results for the non-deformed case:
\begin{equation}
N_{0} =\frac{z}{1-z}\qquad,\qquad
N_{ex} =A \frac{\Gamma(s)}{\beta^s} g_s(z) .  
\end{equation}

The transition temperature is conventionally defined by the condition $N_0 \simeq 0$, implying that nearly all particles are in excited states ($N_{ex} \simeq N$) and the fugacity approaches unity ($z \simeq 1$). For this condition to hold, the number of excited atoms—and thus the generalized function $g_s(z,\nu)$—must remain finite, which requires $\nu > 0$ for any value of $z$. This recovers the usual mathematical constraint for the validity of the deformed algebra, now interpreted as a physical requirement.

Within this framework, the condensation temperature in the Dunkl formalism is readily obtained:
%\begin{equation}
%k_B T_c(\nu)=V_0\left(\frac{2\hbar^2}{ma^2V_0}
%\right)^{\frac{\eta}{\eta+2}}
%\left[\frac{\eta}{2}\frac{\Gamma(D/2)}{\Gamma(D%/\eta)} \frac{N}{g_s(1,\nu)}\right] ^{1/s}, 
%\label{Tc}
%\end{equation}
\begin{equation}
k_B T_c(\nu)=
\left[\frac{N}{A\Gamma(s)g_s(1,\nu)}\right] ^{1/s}, 
\label{Tc}
\end{equation}
which reduces to the standard Bose–Einstein condensation temperature in the absence of deformation. Moreover, the ratio $T_c(\nu)/T_c(1)$ writes 
\begin{equation}
\frac{T_c(\nu)}{T_c(1)}=
\left[\frac{\zeta(s)}{g_s(1,\nu)}\right]^{1/s}. 
\label{ratiotc}
\end{equation}
%showing, in particular, that the dependence on dimensionality and shape of the potential is encoded in the single parameter $s$. 
These expressions provide a natural generalization of the results in \cite{phys5, bm25, phys6,HO24} for the deformed case, as well as those in \cite{gro1, gautam08} for the non-deformed scenario. 

For $s\leq 1$, the divergence of the generalized Bose function $g_s(1,\nu)$ precludes the occurrence of a conventional BEC transition, independent of the deformation parameter $\nu$. To maintain finite values for both the ground-state population $N_0$
Eq.(\ref{ss}) and the excited-state population $N_{ex}$ Eq.(\ref{s}), the fugacity must remain bounded away from unity, ensuring that $N_0$ stays microscopic at all finite temperatures. This behavior is analogous to the well-established absence of true BEC in low-dimensional homogeneous systems~\cite{mermin1966, hohenberg1967} and in certain trapped geometries~\cite{explowerd3, explowerd4, explowerd7}, where quasi-condensation with suppressed density fluctuations but no true long-range order emerges instead. The characterization of quantum correlations and the nature of quasi-condensation within the Dunkl-deformed framework represents an important direction for future investigations.

\begin{figure}[h]
    \centering
    \includegraphics[width=0.5\linewidth]{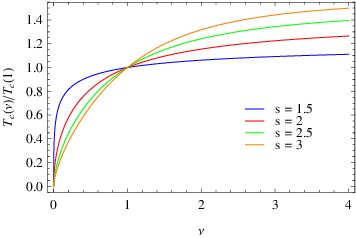}
    \caption{Reduced critical temperature $T_c(\nu)/T_c(1)$ versus \(\nu\) for different values of $s$.}
    \label{fig:ratiotc}
\end{figure}

Figure \ref{fig:ratiotc} shows the ratio (\ref{ratiotc}) as a function of the deformation parameter $\nu$, plotted for various values of the parameter $s$.
%We first note that this ratio decreases as $\nu$ increases. Furthermore, the critical temperature $T_c$ exceeds that of the non-deformed case when $0 < \nu < 1$ (corresponding to negative $\theta$), but falls below it for $\nu > 1$ (corresponding to positive $\theta$). This indicates that the deformation parameter affects the critical BEC temperature asymmetrically.

The critical temperature $T_c(\nu)$ exhibits a systematic dependence on both the universality parameter $s$ and the deformation parameter $\nu$. As $s$ increases, we observe a crossover behavior at $\nu = 1$:
\begin{itemize}
    \item For $\nu < 1$: $T_c(\nu)$ decreases with increasing $s$.
    \item For $\nu > 1$: $T_c(\nu)$ increases with increasing $s$.
\end{itemize}
This crossover behavior manifests in two complementary ways:
\begin{itemize}
\item {\bf Fixed Potential Shape} ($\eta$ constant): When comparing systems with different dimensionalities but the same trap exponent, higher-dimensional systems exhibit lower critical temperatures for $\nu < 1$, but higher critical temperatures for $\nu > 1$. For instance, a 3D system has a lower $T_c$ than its 2D counterpart when $\nu < 1$, but a higher $T_c$ when $\nu > 1$.

\item {\bf Fixed Dimensionality} ($D$ constant): When comparing different trap geometries within the same spatial dimension, steeper confinement potentials (larger $\eta$) lead to higher critical temperatures for $\nu < 1$, but lower critical temperatures for $\nu > 1$. In three dimensions, this means a quartic trap ($\eta = 4$) yields a higher $T_c$ than a harmonic trap ($\eta = 2$) when $\nu < 1$, while the hierarchy reverses for $\nu > 1$.
\end{itemize}
This behavior demonstrates how the Dunkl deformation creates a tunable mechanism for controlling phase transitions through the interplay between confinement geometry and algebraic deformation.

%We observe that, as $s$ increases, the critical temperature $T_c(\nu)$ falls when $\nu < 1$ but rises when $\nu > 1$. This leads to a lower critical temperature in 3D systems compared to 2D for $\nu < 1$, and a higher $T_c$ for $\nu > 1$, assuming a fixed potential shape $\eta$. Conversely, for a fixed dimensionality, $T_c$ increases with increasing $\eta$ when $\nu < 1$, but decreases with $\eta$ when $\nu > 1$. For example, in three dimensions, a harmonic trap yields a lower critical temperature $T_c$ than a quartic trap when $\nu < 1$, whereas the opposite holds true for $\nu > 1$.

%These results highlight how quantum phase transitions can be finely controlled through both algebraic deformation and confinement geometry. For instance, by choosing a deformed system with $\nu > 1$, one can achieve a higher transition temperature, making the phase transition easier to access experimentally.

The condensate fraction $N_0/N$ is also an important parameter to analyze when dealing with Bose-Einstein condensation. Using (\ref{s}) and (\ref{Tc}), we easily get the expression
\begin{equation}
\frac{N_0}{N} = 1 - \left( \frac{T}{T_c(\nu)} \right)^s.
\end{equation}
\begin{figure}[h]
    \centering
    \includegraphics[width=0.5\linewidth]{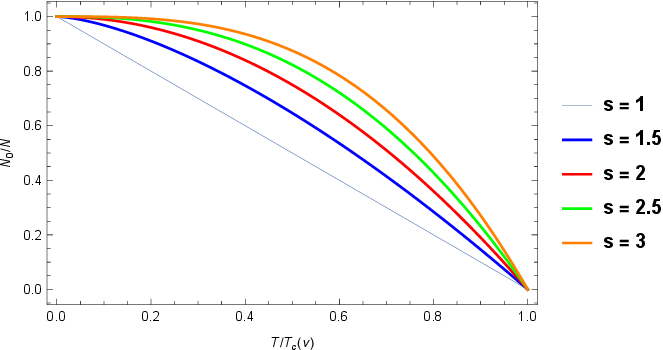}  % Remplace .eps par .pdf si tu utilises pdflatex
    \caption{Condensate fraction $N_0/N$ versus reduced temperature $T/T_c(\nu)$ for various values of $s$.}
    \label{fig:ratioN}
\end{figure}
Figure \ref{fig:ratioN} illustrates the behavior of the condensate fraction \( N_0/N \) as a function of the reduced temperature \( T/T_c \). As expected, the condensate fraction vanishes at the critical temperature and increases monotonically with decreasing temperature, reaching unity at absolute zero. The effect of deformation is hidden in the critical temperature $T_c(\nu)$.

The parameter \( s \), significantly influences the sharpness of the transition. Higher values of \( s \) correspond to a more abrupt onset of condensation, indicating a stronger sensitivity of the condensate formation to temperature changes in such systems. This behavior is consistent with theoretical predictions for Bose-Einstein condensates in systems with varying confining geometries or dimensionality \cite{psw2000, dima2000, ketterle2001, explowerd3, zyl2002, proukakis2003, usama2003, dima2004, had2008, chomaz2015, delfino2017}.

\section{Thermodynamics}
Combining Eqs.~(\ref{uuu}) and (\ref{roeps}), the internal energy is given by
\begin{equation}
U=A\frac{\Gamma(s+1)}{\beta^{s+1}} g_{s+1}(z,\nu), 
\label{u}
\end{equation}
which, when combined with Eq.~(\ref{ampl}), yields the equation of state
\begin{equation}
U=s\,N\,k_B T \left(\frac{T}{T_c (\nu)}\right)^s \frac{g_{s+1}(z,\nu)}{g_{s}(1,\nu)}.
\label{eos}
\end{equation}
This expression reveals significant deviations from the classical equipartition result $U=sNk_BT$. Equation~(\ref{eos}) demonstrates that all thermodynamic properties are determined by only two parameters: the universality parameter $s$ and the deformation parameter $\nu$. This remarkable reduction establishes that diverse physical systems—spanning different spatial dimensions and confinement geometries—exhibit identical thermodynamic behavior when characterized by the same values of $s$ and $\nu$.

To further analyze the phase transition, we compute the heat capacity, $C=\partial U/\partial T|_{z,N}$. This requires distinguishing between two regimes. Below the transition $T<T_c$, one may safely set $z=1$ and therefore,
\begin{equation}
\frac{C_{<}}{Nk_B}=
s(s+1)\frac{g_{s+1}(1,\nu)}{g_s (1,\nu)}\left(\frac{T}{T_c (\nu)}\right)^s.
\label{cvi}
\end{equation}
%The difference with the non-deformed case may %be better visualized by rewriting this %expression as follows:  
%\begin{equation}
%C_{<}(\nu)=
%\frac{1}{2^s}\left(1+\frac{2^s-1}{\nu^s}%\right)C_{<}(\nu=1) ,
%\label{ratio}
%\end{equation}
This result shows that the characteristic $T^s$ dependence of the heat capacity in a degenerate, trapped ideal Bose gas persists despite the deformation of the underlying Heisenberg algebra. For instance, the well-known $T^{D/2}$ and $T^D$ scalings are recovered for a homogeneous or a harmonically trapped gas, respectively, while a $T^{3D/4}$ dependence arises in the case of a quartic trap\cite{phys6, gautam08}.

On the other hand, above $T_c$, $z(T)$ is a quite complicated function in general, but if one recalls that $N_0<<N_{ex}\simeq N$ in this regime, the equation (\ref{s}) simplifies drastically to yield a simpler implicit relation between $z$ and $T$:
\begin{equation}
\frac{g_s(z,\nu)}{g_s(1,\nu)}=\left(\frac{T_c (\nu)}{T}\right)^s ,
\label{z}
\end{equation} 
which allows obtaining the following final formula for $C_>$:
\begin{equation}
\frac{C_{{>}}}{Nk_{B}}=s(s+1)\frac{
g_{s+1}(z,\nu)}{g_s (z,\nu) }-s^2\frac{
g_{s}(z,\nu )}{g_{s-1} (z,\nu) }.
\label{cvs}
\end{equation}
It is important to highlight that the expressions (\ref{cvi}) and (\ref{cvs}) for $C_<$ and $C_>$ constitute natural generalizations of the results reported in \cite{gro1,gautam08,bm25,phys6, HO24, phys5}. They unify both deformed and non-deformed scenarios, extending the analysis to arbitrary spatial dimensions and encompassing all power-law traps, including the limiting case of an untrapped gas.

Furthermore, although not immediately apparent, the overall temperature dependence of $C_>$  remains unaffected by the deformation parameter. To make this explicit, let us examine the behavior near the critical point $T\to T_c^{+}$ ($z\to 1^{-}$). From Eq. (\ref{z}), we readily obtain:
\begin{equation}
1-z\simeq \frac{(1-\nu^{1-s})(1-2^{1-s})}{(1-\nu^{2-s})(1-2^{2-s})}\frac{\zeta(s)}{\zeta(s-1)}-
 \frac{1}{(1-\nu^{2-s})(1-2^{2-s})\zeta(s-1)}\left(\frac{T_c}{T}\right)^s
\label{zcrit}
\end{equation}
Expanding the generalized Bose functions in Eq. (\ref{cvs}) around $z=1$, and employing Eq. (\ref{zcrit}), we readily obtain the following expression, where $a$ and $b$ are functions of $\nu$ and $s$ alone (their explicit forms are not relevant for the present discussion):
\begin{equation}
\frac{C_>(\nu)}{Nk_B}=a(\nu,s)+b(\nu,s)\left(\frac{T_c}{T}\right)^{s}.
\label{cvcrit}
\end{equation}
This result unambiguously indicates that the deformation does not alter the characteristic $T^{-s}$ critical behavior of the heat capacity.

\begin{figure}[h]
    \centering
    \includegraphics[width=0.7\linewidth]{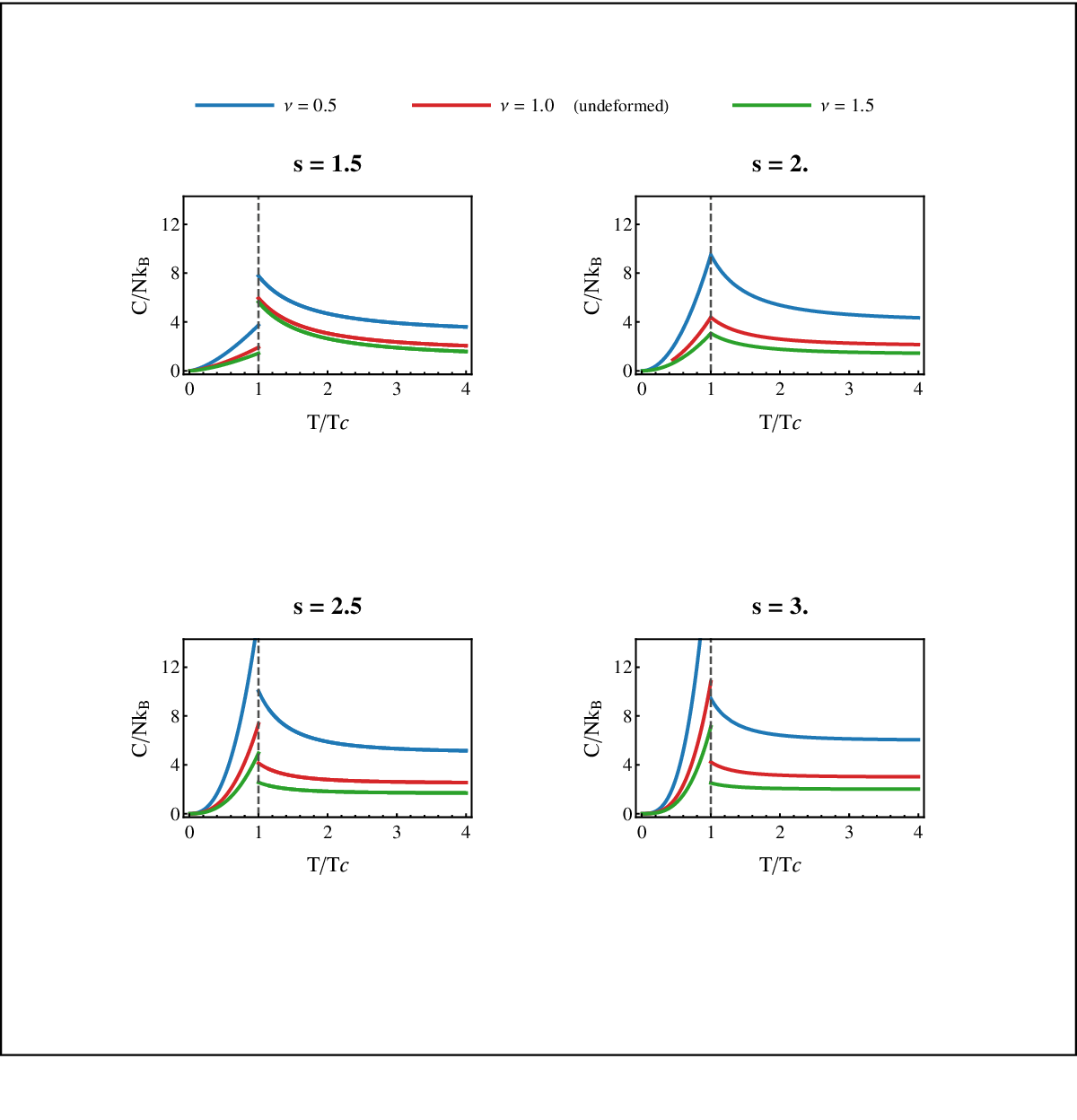}
    \caption{Heat capacity (normalized to $Nk_B$) as a function of reduced temperature $T/T_c(\nu)$ for various universality parameter values $s$. The four panels show $s = 1.5$ (upper left), $s = 2.0$ (upper right), $s = 2.5$ (lower left), and $s = 3.0$ (lower right). Three deformation parameters are displayed: $\nu = 0.5$ (blue curves), $\nu = 1.0$ (red curves, undeformed case), and $\nu = 1.5$ (green curves). The vertical dashed line marks the critical temperature $T_c$. Note that $s = 2.0$ exhibits a continuous transition without discontinuity, while all other $s$ values show characteristic jumps at the phase transition.}
    \label{fig:figure2}
\end{figure}

Figure \ref{fig:figure2} shows the heat capacity across the full range of reduced temperature $T/T_c$ for four values of the parameter $s$: $s=3/2$ (upper left), $s=2$ (upper right), $s=5/2$ (lower left), and $s=3$ (lower right). Each panel displays curves for three values of the deformation parameter: $\nu=0.5$ (blue), $\nu=1$ (red: undeformed case), and $\nu=1.5$ (green). 
The characteristic $\lambda-$point behavior is evident across all investigated values of $s$ and $\nu$. In general, the heat capacity exhibits a discontinuity at the critical temperature, signaling a second-order phase transition. A notable exception occurs at $s=2$, where the heat capacity instead displays a sharp but continuous peak at the transition point. This distinction originates from Eq. (\ref{cvs}), where the second term involves the function $g_1(z, \nu)$, which diverges as $z\to 1$. The absence of a discontinuity at $s=2$ was previously reported in \cite{bm25} for the specific case of the two-dimensional harmonic oscillator. In the present framework, this behavior characterizes the universality class defined by $s=2$, encompassing, for instance, a $x^{2/3}$ potential in one dimension and a $x^6$ potential in three dimensions. In such systems, the phase transition remains continuous, though it features a pronounced peak in the heat capacity—reminiscent of a Berezinskii–Kosterlitz–Thouless (BKT) transition. Finally, we would like to highlight the significant influence of the deformation parameter. For $0<\nu < 1$, the heat capacity exceeds that of the undeformed case, indicating enhanced energy fluctuations. Conversely, for $\nu > 1$, the heat capacity is suppressed, reflecting reduced fluctuations under stronger deformations.

To gain deeper insight into the behavior of the heat capacity at the critical point, we examine the discontinuity at $T_c$, given by:
\begin{equation}
\Delta C=\frac{C_{<}-C_{>}}{Nk_{B}}|_{T_c}=
\frac{\zeta(s)}{\zeta(s-1)}\frac{s^2}{2\nu}\frac{\nu^{s-1}+2^{s-1}-1}{\nu^{s-2}+2^{s-2}-1}.
\label{jump}
\end{equation}
We observe that for $\nu=1$, the discontinuity $\Delta C$ reduces to the standard expression $s^2\zeta(s)/\zeta(s-1)$. In particular, this recovers the known result $\Delta C=54\zeta(3)/\pi^2$ for the three-dimensional harmonic oscillator ($s=3$), as reported in \cite{bm25, HO24}. Moreover, irrespective of the deformation, $\Delta C=0$ when $s=2$, in agreement with the earlier discussion.

\begin{figure}[h]
    \centering
    \includegraphics[width=0.6\linewidth]{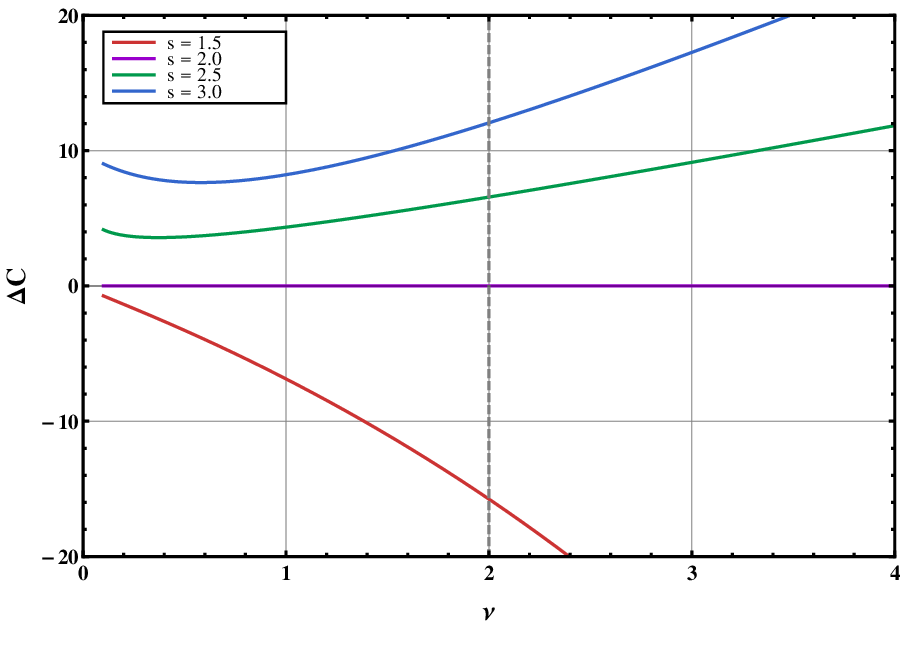}
    \caption{Discontinuity $\Delta C$ defined in equation (\ref{jump}) versus the deformation parameter $\nu$ for four universality parameters. $s=1.5$ (red): $\Delta C<0$,
    $s=2$ (purple): Continuous BKT-like transition, $s=2.5$ (green) and $s=3$ (blue): $\Delta C>0$. The vertical dashed line at $\nu=2$ depicts the critical boundary for physical validity of the Dunkl formalism (see section 4).}
    \label{fig:figure5}
\end{figure}

Figure~\ref{fig:figure5} shows the heat capacity discontinuity [Eq.~(\ref{jump})] as a function of both deformation and universality parameters. The phase transition exhibits conventional behavior with positive heat capacity jumps ($\Delta C > 0$) for $s > 2$, but becomes anomalous with negative jumps ($\Delta C < 0$) in the range $1 < s < 2$. This anomalous regime encompasses $2/3 < \eta < 2$ (one dimension), $\eta > 2$ (two dimensions), and $\eta > 6$ or $\eta < -6$ (three dimensions). The case $s = 3/2$ illustrated includes physically distinct systems such as one-dimensional linear traps and two-dimensional quartic confinement.
The negative discontinuity reveals anomalous thermodynamics: heat capacity decreases upon condensation, indicating enhanced thermal stability below $T_c$. Remarkably, this behavior is independent of the deformation parameter $\nu$, thus characterizing entire universality classes rather than specific trap geometries.

Experimentally, three-dimensional inverse power-law potentials ($\eta < -6$) represent the most accessible route to observe negative heat capacity jumps. Such attractive potentials—realizable through modified Coulomb interactions, long-range molecular forces, or engineered optical traps—are more experimentally feasible than the steep repulsive potentials ($\eta > 6$) required in the same dimension. This accessibility significantly expands the parameter space for experimental verification.

%Importantly, the regime $s<1$ encompasses potentials that resemble interatomic van der Waals interactions. However, such a potential on its own is insufficient to induce Bose-Einstein condensation (BEC) in three dimensions. To enable BEC, an additional short-range repulsive term, typically $r^{-12}$, must be included—resulting in the well-known Lennard-Jones potential.

\section{Classical regime}

The high-temperature regime ($T \gg T_c$) corresponds to $g_s(z,\nu) \ll 1$ [Eq.~(\ref{z})], which requires $z \ll 1$. Expanding the generalized Bose functions for small $z$ yields
\begin{equation}\label{bosef}
g_s(z,\nu) \simeq \frac{z^2}{2^{s-1}} + \frac{z^\nu}{\nu^{s-1}} + \mathcal{O}(z^{2\nu}),
\end{equation}
where the leading-order behavior depends on $\nu$.

For $0 < \nu < 2$, the heat capacity becomes
\begin{equation}
C_> = \frac{s}{\nu}Nk_B.
\end{equation}
This result confirms enhanced heat capacity for $0 < \nu < 1$ and suppression for $\nu > 1$ relative to the undeformed case ($\nu = 1$).

At the critical value $\nu = 2$, both $z^2$ and $z^\nu$ terms contribute equally in Eq.~(\ref{bosef}), yielding
\begin{equation}\label{classical}
C_> = s Nk_B,
\end{equation}
which exactly reproduces the undeformed classical result. For $\nu > 2$, the $z^2$ term dominates, giving
\begin{equation}
C_> = \frac{s}{2}Nk_B.
\end{equation}
This $\nu$-independent result differs from the undeformed case, indicating thermodynamically inconsistent behavior. Consequently, physical consistency requires $0 < \nu \leq 2$.

This constraint generalizes previous results~\cite{bm25} to arbitrary power-law traps in any spatial dimension. Unlike earlier studies treating $\nu$ as a free parameter with only a lower bound, our analysis establishes universal upper and lower bounds. Since the limits $\eta \to \pm\infty$ yield $s \to D/2$, any regular potential admitting a power series expansion must satisfy the same constraint. Therefore, the Dunkl deformation's validity is fundamentally restricted: classical consistency imposes the universal bound $0 < \nu \leq 2$, independent of trap geometry or spatial dimension.

\section{Conclusions}
We have investigated the thermodynamic properties of a $D$-dimensional ideal Bose gas in general power-law potentials within the Dunkl-deformed formalism. Our analysis extends previous results for harmonic confinement to arbitrary power-law traps, encompassing homogeneous systems as a limiting case.

The central finding is the emergence of a universal parameter $s = D(1/\eta + 1/2)$ that completely determines the system's thermodynamic behavior. This universality implies that physically distinct systems—differing in spatial dimension $D$ and confinement exponent $\eta$—exhibit identical thermodynamics when characterized by the same $s$ value. Bose-Einstein condensation occurs exclusively for $s > 1$, consistent with the Mermin-Wagner-Hohenberg theorem. The heat capacity displays characteristic power-law scaling: $T^s$ dependence below $T_c$ and $T^{-s}$ behavior near the critical point. The BEC transition remains second-order for all $s \neq 2$, while $s = 2$ yields continuous transitions reminiscent of Berezinskii-Kosterlitz-Thouless physics.

Critically, we establish universal bounds on the Wigner parameter: $0 < \nu \leq 2$. This constraint, derived from classical limit consistency, applies to any power-law potential in arbitrary dimensions. For $\nu > 2$, the formalism produces unphysical thermodynamics, including anomalous heat capacity scaling $C \propto s/2$, corresponding to effectively halved degrees of freedom. These bounds extend to all regular potentials admitting power series expansions, demonstrating that previous treatments of $\nu$ as a free parameter are fundamentally incorrect.

Our results provide definitive physical boundaries for Dunkl-deformed quantum statistics, establishing when such algebraic deformations yield meaningful descriptions of quantum many-body systems. The universal nature of these constraints underscores their fundamental importance for future applications of the Dunkl formalism in quantum statistical mechanics.

%As a direction for future work, it would be valuable to explore how particle interactions and trap geometry could influence the upper bound on the Wigner parameter. Investigations in this direction are currently underway and will be reported in forthcoming publications.

%Beyond equilibrium systems, future research may also consider the role of the Wigner parameter in non-equilibrium or time-dependent settings. These avenues could provide further insights into the broader applicability of the Dunkl formalism to quantum gases.

\section*{Acknowledgments}
%{\color{red} The authors thank the anonymous reviewer for %constructive comments.} 
This work is supported by the Ministry of Higher Education and Scientific Research, Algeria under the code: PRFU:B00L02UN020120220002. 
%B. C. L\"{u}tf\"{u}o\u{g}lu is supported by the Internal Project %[2023/2211], of Excellent Research of the Faculty of Science of %Hradec Kr\'{a}lov\'{e} University.

\section*{Data Availability Statements}

The authors declare that the data supporting the findings of this study are available within the article.

%\section*{Declaration of Generative AI and AI-assisted technologies in the writing process}

%The authors did not use any generative AI and AI-assisted technologies in the writing process.

\end{document}